\documentclass {llncs}
\usepackage{amssymb}
\usepackage{epsfig}
\usepackage{epsfig}

\usepackage{pifont}% http://ctan.org/pkg/pifont
\newcommand{\cmark}{\ding{51}}%
\usepackage[hyphens]{url}
\usepackage{hyperref}

\PassOptionsToPackage{hyphens}{url}\usepackage{hyperref}

\makeatletter
\def\set@curr@file#1{%
  \begingroup
    \escapechar\m@ne
    \xdef\@curr@file{\expandafter\string\csname #1\endcsname}%
  \endgroup
}
\def\quote@name#1{"\quote@@name#1\@gobble""}
\def\quote@@name#1"{#1\quote@@name}
\def\unquote@name#1{\quote@@name#1\@gobble"}
\makeatother

\usepackage{amssymb}
\usepackage{acronym}
 \usepackage{url}
 \usepackage{multirow}
 \usepackage[all]{xy}
 \usepackage{amsmath}
 \usepackage{float}
  \usepackage{amssymb,amsmath,xspace,graphicx,multirow}
\restylefloat{table}
 
\usepackage{multirow}
\usepackage[all]{xy}
\usepackage{amsmath}
\urldef{\mailsc}\path|adnan.rashid@seecs.nust.edu.pk|
\urldef{\mailsb}\path|siddiu3@mcmaster.ca|
\urldef{\mailsa}\path|{muh_sidd,tahar}@ece.concordia.ca|
\begin{document}

\pagestyle{empty}

\mainmatter

\title{Formal Verification of Cyber-Physical Systems using Theorem Proving}

\titlerunning{Formal Verification of Cyber-Physical Systems}

\author{Adnan Rashid$^1$\and Umair Siddique$^2$\and Sofiène Tahar$^2$}

\authorrunning{A. Rashid, U. Siddique and S. Tahar}

\institute{$^1$School of Electrical Engineering and Computer Science (SEECS)\\
National University of Sciences and Technology (NUST)\\
Islamabad, Pakistan\\
\mailsc\\
$^2$Department of Electrical and Computer Engineering \\
Concordia University, Montreal, Canada \\
\mailsa\\ }

\maketitle

\begin{abstract}
Due to major breakthroughs in software and engineering technologies, embedded systems are increasingly being utilized in areas ranging from aerospace and next-generation transportation systems, to smart grid and smart cities, to health care systems, and broadly speaking to what is known as Cyber-Physical Systems (CPS). A CPS is primarily composed  of  several  electronic,  communication  and  controller  modules and some actuators and sensors. The mix of heterogeneous underlying smart technologies poses a number of technical challenges to the design and more severely to the verification of such complex infrastructure. In fact, a CPS shall adhere to strict safety, reliability, performance and security requirements, where one needs to capture both physical and random aspects of the various CPS modules and then analyze their inter-relationship across interlinked continuous and discrete dynamics. Often-times  however,  system  bugs  remain  uncaught  during  the  analysis and in turn cause unwanted scenarios that may have serious consequences in safety-critical applications. In this paper, we introduce some of the challenges surrounding the design and verification of contemporary CPS with the advent of smart technologies. In particular, we survey recent developments in the use of theorem proving, a formal method, for the modeling, analysis and verification of CPS, and overview some real world CPS case studies from the automotive, avionics and healthtech domains from system level to physical components.
\end{abstract}

\section{Introduction}

\begin{figure}[!ht]
\centering
\scalebox{0.400}
{\hspace*{-0.26cm}\includegraphics[trim={0.0 0.0cm 0.0 0.0cm},clip]{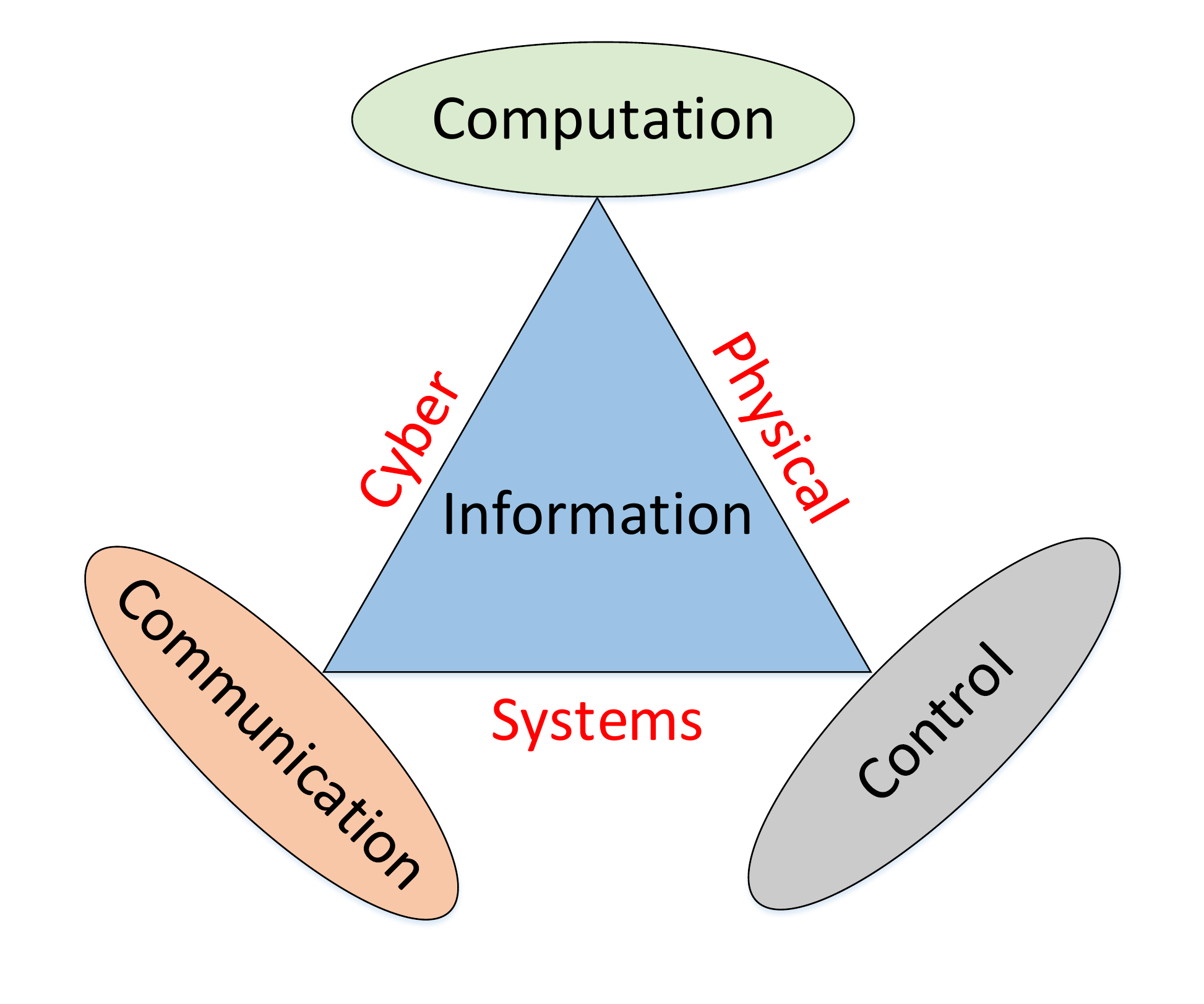}}
\caption{Components of a CPS~\cite{cps2020figure}}
\label{FIG:cps_components}
\end{figure}

Cyber-Physical systems (CPS)~\cite{rajkumar2010cyber} are engineered systems involving a cyber component that controls the physical components, as shown in Figure~\ref{FIG:cps_components}. The cyber elements include embedded systems and network controllers, which are usually modeled as discrete events. Whereas, the physical components exhibit continuous dynamics, such as the physical motion of a robot in space or the working of an analog circuit, and are commonly modeled using differential equations.
CPS are capable of performing two main functionalities (a) constructing the cyber space using intelligent data management, computational and analytical capabilities; and (b) real-time data acquisition from the physical world and information feedback from the cyber space using some advanced connectivity, as depicted in Figure~\ref{FIG:cps_components}. They can be small, such as artificial pancreas, or very large and complex, such as a smart car or smart energy grid.
The development of powerful embedded system hardware, low-power sensing and widely deployed communication networks has drastically increased the dependence of system functionality on CPS.
CPS are widely used in advanced automotive systems (autonomous vehicles and smart cars), avionics, medical systems and devices, optical systems, industrial process control, smart grids, traffic safety and control, robotics and telecommunication networks, etc. For example, smart (self-driving) cars are considered as a highly complex autonomous CPS composed of over one hundred processors, and an array of sensors and actuators that interact with the external environment, like the road infrastructure and internet.

The main goals for an efficient design of CPS are to co-design its cyber and physical parts, and to engineer the system of systems involving the intrinsic heterogeneity. Moreover, an increase in the complexity of its various components and the utilization of advanced technologies pose a major challenge for developing a CPS. For example, in the case of smart cars, it is required to develop cost-effective methods ensuring: a) design and analysis (verification) of its various components at different levels of abstraction, i.e., at different systems and software architecture levels; b) analyzing and understanding the interactions of system of systems, e.g., cars' control system and its various components, such as engine, wheel, steering; c) minimizing the cost of the car by ensuring the safety, reliability, performance and stability of the overall system. Thus, these requirements have to be fulfilled for the efficient design and analysis of a CPS.

The analysis of CPS can generally be characterised as of three types, namely, functional, performance and dependability analysis. For example, the functional analysis involves the analysis of the physical, control and signal processing components of CPS. Each of these characteristics also need to consider a hybrid behavior incorporating both continuous and discrete dynamics, e.g., the physical and cyber elements of the underlying system.

Conventionally, CPS are analyzed using paper-and-pencil methods or computer-based numerical and symbolic techniques. Moreover, most of the time is spent on designing the life-cycle of CPS and their physical (dynamical) behaviour needs to be manipulated. However, there is a lack of theoretical foundations for CPS dynamics and compositional theories for the heterogeneous systems in the tools associated with these analyses. Moreover, these analysis methods suffer from their inherent limitations, like human-error proneness, discretization and numerical errors and the usage of unverified simplification algorithms~\cite{duran2013misfortunes} and thus cannot provide absolute accuracy of the corresponding analysis. Due to the safety critical-nature of CPS, the accuracy of their design and analysis is becoming a dire need. For example, the fatal crash of Uber's self-driving car in March $2018$ that killed a pedestrian in Tempe, Arizona, USA was found to be caused by some sensor's anomalies~\cite{uber2018crash}. A more rigourous analysis of CPS could have avoided this incident.

Formal methods~\cite{hasan2015formal} have been used as a complementary technique for analyzing CPS and thus can overcome the above-mentioned inaccuracy limitations of the analysis. The two most commonly used formal methods are model checking~\cite{baier2008principles} and theorem proving~\cite{harrison2009handbook}. Model checking is based on developing a state-space based model of the underlying system and formally verifying the properties of interest, specified in temporal logic. It has been used for analyzing several aspects of a CPS~\cite{clarke2011statistical}. However, this kind of analysis involves the discretization of the continuous dynamical models and thus compromises the accuracy of the corresponding analysis. Moreover, it also suffers from the state-space explosion problem~\cite{baier2008principles}.
Theorem proving~\cite{harrison2009handbook} is a computer based mathematical method that involves developing a mathematical model of the given system in an appropriate logic and the formal verification of the properties of interest based on mathematical reasoning within the sound core of a theorem prover.
The involvement of the formal model and its associated formally specified properties along with the sound nature of theorem proving ensures the accuracy and completeness of the analysis.
Based on the decidability or undecidability of the underlying logic, e.g., propositional or higher-order logic, theorem proving can be automatic or interactive, respectively.

Many theorem provers, e.g., HOL4~\cite{norris_hol}, HOL Light~\cite{harrison1996hol}, Isabelle~\cite{paulson1994isabelle}, KeYmaera~\cite{platzer2008keymaera}, Coq~\cite{bertot2013interactive}, PVS~\cite{owre1992pvs} have been used for the formal analysis (formal verification) of CPS, e,g., formal functional analysis, formal probabilistic and performance analysis, formal dependability analysis, and hybrid analysis. For instance, the KeYmaera theorem prover has been specifically designed for the formal verification of hybrid systems, thus, incorporating both the continuous and discrete dynamics of the underlying system. KeYmaera is based on deductive reasoning and computer algebraic prover technologies. It uses differential dynamic logic for the model implementation and specification of the underlying system, which is a first-order logic. Similarly, HOL Light provides an extensive support of mathematical libraries that have been used for the functional analysis, i.e., the verification of various continuous aspects of CPS, such as control systems, power electronics, electromagnetic, quantum and optical systems. HOL4 and Isabelle theorem provers provide an extensive support for the formal probabilistic and dependability analysis of systems. Likewise, Isabelle and HOL4 have been extensively used for the verification of software components, providing safety and security analysis of the underlying CPS.
In this paper, we report these developments that have been done for the modeling, analysis and verification of CPS in these theorem provers.

\section{Formal Functional Analysis}\label{SEC:Formal_functional_analysis}

\subsection{Verification of Physical Components}

Hasan et al.~\cite{hasan2009formalow} proposed a framework for analyzing the optical waveguides using HOL4. In particular, the authors formally analyzed the eigenvalues for the planar optical waveguides and utilized their proposed framework for analyzing a planar asymmetric waveguide.
Afshar et al.~\cite{afshar2014formalization} developed a formal support for the complex vector analysis using HOL Light and used it to formally verify the law of reflection for the planar waves.
Later, the authors used the formalization of complex vectors to formalize the notions of electromagnetic optics~\cite{khan2014formalsp}, which is further used for performing the formal analysis of the resonant cavity enhanced photonic devices.

Siddique et al.~\cite{siddique2014framework} provided a formalization of geometrical optics using HOL Light. The authors formalized fundamental concepts about geometrical optics, i.e., ray, free space, optical system and its stability. Finally, they used their proposed formalization to perform the stability analysis of the Fabry-Perot resonator with fiber rod lens~\cite{siddique2013formal}. Next, the authors extended their framework by formalizing the ray optics of the cardinal points and utilized it for formally analyzing a thick lens~\cite{siddique2014towards2} and the optical instrument used to compensate the ametropia of an eye~\cite{siddique2015formalization}. Moroever, the authors formalized the notion of optical resonators and used it for formally verifying the 2-D microresonator lattice optical filters~\cite{siddique2014towards}. Finally, the authors extended their formal support for geometrical optics in HOL Light by performing the formal analysis of the gaussian~\cite{siddique2016formal} and periodic~\cite{siddique2017formal} optical systems.

As a part of the optics formal verification project~\cite{khan2014formal}, Mahmoud et al.~\cite{mahmoud2013formalization} provided a support for the formal analysis of the quantum systems using HOL Light. In particular, the authors formalized the infinite dimension linear spaces and used it for formally verifying a quantum beam splitter. Next, the authors used their formalization of linear algebra to formalize the optical quantum circuits, i.e., the flip gate and used it to formally verify the beam splitter and the phase conjugating mirror~\cite{mahmoud2014formal}. Later, the authors also formalized the notion of coherent light, which is a light produced by the laser sources and formally verified its various properties using HOL Light~\cite{mahmoud2014quantum}.
Based on these findings, Beillahi et al.~\cite{beillahi2016hierarchical} proposed a framework for the hierarchical verification of the quantum circuit and used it for the formal analysis of a controlled-phase gate and the Shor’s factoring quantum circuits.
Rand et al.~\cite{rand2018qwire} proposed a framework implementing the QWIRE quantum circuit language in Coq, which accepts a high-level abstract model of the quantum circuits and allows the verification of their properties using Coq's features such as dependently-typed circuits and proof-carrying code. Liu et al.~\cite{liu2019formal} formalized the theory of Quantum Hoare Logic (QHL) and used it for formally verifying the correctness of a nontrivial quantum algorithm using Isabelle.

\subsection{Verification of Software Components}

The High-Assurance Cyber Military Systems (HACMS) research program~\cite{fisher2017hacms} was started by the Defense Advanced Research Projects Agency (DARPA) in the USA with an aim of creating a technology for constructing CPS that are resilient against cyber-attacks, i.e., CPS providing an appropriate security and safety properties. One of the major goals of this program is to create a high-assurance software for vehicles, ranging from automobiles to military vehicles, such as quadcopters and helicopters. As a part of this project, Cofer et al.~\cite{cofer2018formal} proposed a formal approach for constructing a secure airvehicle software to ensure security against cyber attacks using Isabelle. Moreover, the authors applied their proposed approach for formally analyzing the SMACCMcopter, which is a modified commercial quadcopter, and Boeing’s Unmanned Little Bird (ULB), which is a full-sized optionally-piloted helicopter.
Klein et al.~\cite{klein2009sel4} presented the formal verification of seL4 microkernel in HOL4, which is a third-generation microkernel of L4 provenance. The authors formally proved that the implementation of the underlying system follows the high-level specification of the kernel behaviour using Isabelle. Moreover, they also verified two vital properties of the microkernel, i.e., 1) the kernel will not perform an unsafe operation; 2) it will never crash.

\subsection{Verification of Control and Signal Processing Components}

Transform methods, such as Laplace, Fourier and $z$-transforms are widely used for solving dynamical models and performing the frequency domain analysis of systems. Generally, the dynamics of a system in frequency domain are characterized by the transfer function and frequency response, providing a relationship between its input and output and are important properties of the control and signal processing components of a CPS. In this regards, Taqdees et al.~\cite{taqdees2013formalization} formalized the Laplace transform using multivariate calculus theories of HOL Light. Moreover, the authors used their formalization of the Laplace transform for formally verifying the transfer function of the Linear Transfer Converter (LTC) circuit.
Next, the authors extended their framework and provided a support to formally reason about the linear analog circuits, such as Sallen-Key low-pass filters~\cite{taqdees2017tflac} by formalizing the system governing laws such as Kirchhoff's Current Law (KCL) and Kirchhoff's Voltage Law (KVL) using HOL Light. Later, Rashid et al.~\cite{rashid2017tmformalization} proposed a new formalization of the Laplace transform based on the notion of sets and used it for analyzing the control system of the Unmanned Free-swimming Submersible (UFSS) vehicle~\cite{rashid2017formal} and 4-$\pi$ soft error crosstalk model~\cite{adnan2018JAL}. The Laplace transform~\cite{laplace2019isabelle,wang2017formalization} has also been formalized in Isabelle and Coq theorem provers. Similarly, Rashid et al.~\cite{rashid2016formalization} formalized the Fourier transform in HOL Light and used it to formally analyze an Automobile Suspension System (ASS), an audio equalizer, a drug therapy model and a MEMs accelerometer~\cite{rashid2017formalft}.

To perform the transfer function based analysis of the discrete-time systems, Siddique et al.~\cite{siddique2014formalization} formalized $z$-transform using HOL Light and used it for the formal analysis of Infinite Impulse Response (IIR) Digital Signal Processing (DSP) filter. Later, the authors extended their proposed framework by providing the formal support for the inverse $z$-transform and used it for formally analyzing a switched-capacitor interleaved DC-DC voltage doubler~\cite{siddique2018formal}. Beillahi et al.~\cite{beillahi2015formal} proposed a formalization of signal-flow graph, which is widely used for evaluating the system performance in the form of transfer function, using HOL Light. The authors used their proposed framework for formally analyzing a die design process~\cite{beillahi2014towards}, $1$-boost cell interleaved DC-DC, Pulse Width Modulation (PWM) push-pull DC-DC converters~\cite{beillahi2015formal}, Double-coupler Double-ring (DCDR) photonic processor~\cite{siddique2015formal}, z-source impedance network and PANDA Vernier resonator~\cite{beillahi2016formal}.

Farooq et al.~\cite{farooq2013formal} proposed a formal framework for the kinematic analysis of a two-link planar manipulator, which describes a geometrical relationship between the robotic joints and links, and is widely used to capture the motion of the robots. Moreover, the authors performed the formal kinematic analysis of a biped walking robot using HOL Light. Next, Affeldt et al.~\cite{affeldt2017formal} carried forward this idea and formalized the foundational support for 3D analysis of the robotic manipulators in Coq. The authors used their proposed framework for the kinematic analysis of the SCARA robot manipulator. Wu et al.~\cite{wu2017formalization} used HOL4 to formally reason about the forward kinematics of the 3-DOF planar robot manipulator. Similarly, Li et al.~\cite{li2014formal} provided the formal verification of the Collision-free Motion Planning Algorithm (CFMPA) of Dual-arm Robot (DAR) using HOL4. Walter et al.~\cite{walter2010experiences} formally verified a collision-avoidance algorithm for service robots in Isabelle. The authors mainly formalized the safety zone of the robot based on the algorithm and used it to formally verify that the robot will stop upon facing an obstacle, otherwise, it will continue its movement within the safety zone. Recently, Rashid et al.~\cite{rashid2018formalrobotics} provided the formal modeling and analysis of the $2$-DOF robotic cell injection systems using HOL Light.

\subsection{Formal Hybrid Analysis}

Platzer et al.~\cite{platzer2009computing} developed an algorithm for the verification of the safety properties of CPS. The authors used the notion of continuous generalization of induction to compute the differential invariants, which do not require solving the differential equations capturing the dynamics of CPS. Moreover, they used their proposed algorithm for formally verifying the collision avoidance properties in car controls and aircraft roundabout maneuvers~\cite{platzer2009formal} using KeYmaera.
Similarly, Platzer et al.~\cite{platzer2009european} verified the safety, controllability, liveness, and reactivity properties of the European Train Control System (ETCS) protocol using KeYmaera.
KeYmaera has also been widely used for the dynamical analysis of various CPS, such as a distributed car control system~\cite{loos2011adaptive}, freeway traffic control~\cite{mitsch2012towards}, autonomous robotic vehicles~\cite{mitsch2013provably} and industrial airborne
collision avoidance system~\cite{jeannin2015formal}.
Recently, Bohrer et al.~\cite{bohrer2018veriphy} presented VeriPhy, a verified pipeline for automatically transforming verified models of CPS to verified controller executables. It proves CPS safety at runtime by verified monitors.
All these analysis performed using KeYmaera are based on the differential dynamics logic, which captures both the continuous and discrete dynamics of CPS and their interaction. This logic allows the suitable automation of the verification process as well. Similarly, Foster et al.~\cite{foster2017towards} proposed a framework for the verification of CPS based on Unifying Theories of Programming (UTP) and Isabelle/HOL. In particular, the authors provide the implementation of designs, reactive processes, and the hybrid relational calculus, which are important foundational theories for analyzing CPS.

\section{Formal Probabilistic and Performance Analysis}

Hasan et al.~\cite{hasan2015formalized} proposed a higher-order logic framework for the probabilistic analysis of the systems using HOL4. The authors first formalized the standard uniform random variable~\cite{hasan2007formalization}. Next, they used this random variable alongside a non-uniform random number generation method to formalize continuous uniform random variables. Finally, the authors used their proposed formalization for the probabilistic analysis of roundoff error in a digital processor~\cite{hasan2007formalization}.
Next, Hasan et al.~\cite{hasan2008using} used HOL4 for the formal verification of the expectation and variance of the discrete random variable and used their expectation theory to formally reason about the Coupon Collector’s problem~\cite{hasan2008using}.
Later, the authors extended their framework by providing the formal verification of the expectation properties of the continuous random variables, i.e., Uniform, Triangular and Exponential~\cite{hasan2009formalcont}.
Next, the authors formalized the indicator random variables using HOL4 and used it for the expected time complexity analysis of various algorithms, i.e., the birthday paradox, the hat-check and the hiring problems~\cite{hasan2010formally}.
Elleuch et al.~\cite{elleuch2015formal} used the probability theory of HOL4 to formally reason about the detection properties of Wireless Sensor Networks (WSNs) and a WSN-based monitoring framework~\cite{elleuch2016formal}. Moreover, the authors conducted the performance analysis of WSNs~\cite{elleuch2013towards}.
Hasan et al. also used their probability theory in HOL4 for conducting the performance analysis of Automatic-repeat-request (ARQ) protocols, i.e., Stop-and-Wait, Go-Back-N and Selective-Repeat protocols~\cite{hasan2008performance}.
Finally, Hasan et al.~\cite{hasan2011reasoning} formalized the notion of conditional probability and formally verified its classical properties, i.e., Bayes' theorem and total probability law. The authors utilized their formalization for formally analyzing the binary asymmetric channel, which is widely used in communication systems.
Mhamdi et al.~\cite{mhamdi2010formalization} formalized the Lebesgue integral using HOL4 and used it for formally verifying the Markov and Chebyshev inequalities, and the Weak Law of Large Numbers (WLLN) theorem.
Next, the authors built upon Lebesgue integral to formalize the Radon-Nikodym derivative and used it for formalizing the fundamentals of information theory, i.e., Shannon and relative entropies~\cite{mhamdi2011formalization}.
Later, Mhamdi et al.~\cite{mhamdi2015evaluation} used the probabilistic analysis support developed in HOL4 to evaluate the security properties of the confidentiality protocols.
%%%%%%%%%%%%%%%%%%%%%%%%%%%%%%%%%%%%%%%%%%%%%%%%%%%%%%%%%%%%%%%%%%%%%%%%%%%%%%%%%%%%%%%%%%%%%%%%%%%%%%%%%%%%%%%%%%%%%
A library for the formal probabilistic analysis has also been developed in Isabelle. Holzl et al.~\cite{holzl2011three} formalized measure theory with extended real numbers as measure values, in particular, the authors formalized Lebesgue integral, product measures and Fubini’s theorem using Isabelle. Eberl at al.~\cite{eberl2015verified} developed an inductive compiler, which takes programs in a probabilistic functional language and computes density functions for the probability spaces using Isabelle. Similarly, Holzl et al.~\cite{holzl2012interactive} proposed a formalization of Markov chains and used it to formally verify the ZeroConf and  the Crowds protocols using Isabelle.

\section{Formal Dependability Analysis}\label{SEC:formal_dependability_analysis}

Hasan et al.~\cite{hasan2010formal} formalized some fundamental concepts about the reliability theory in HOL4 and used it for formal reliability analysis of reconfigurable memory arrays in the presence of stuck-at and coupling faults. Moreover, the authors performed the reliability analysis of the combinational circuits, such as full adders, comparators and multiplier.
Later, Abbasi et al.~\cite{abbasi2014approach} extended the reliability analysis framework by formally verifying some statistical properties, i.e., second moment and variance and other reliability concepts, i.e., survival, hazard and fractile functions. The authors utilized their proposed framework for formally analyzing the essential electronic and electrical system components.

Liu et al.~\cite{liu2011formalization} proposed a framework to reason about the finite-state discrete-time Markov chains using HOL4 and formally verified some of its properties such as joint and steady-state probabilities, and reversibility. The authors utilized their proposed framework to formally analyze a binary communication channel and an automatic mail quality measurement protocol~\cite{liu2013formaljour}. Next, the authors formalized the discrete-time Markov reward models and used it to formally reason about the memory contention problem of a multi-processor system~\cite{liu2013formal}.
Later, the authors proposed a framework to formally reason about the properties of the Hidden Markov Models (HMMs) such as joint probabilities and formally analyzed a DNA sequence~\cite{liu2014formal}.

Ahmad et al.~\cite{ahmad2016formal} developed a higher-order logic based framework for the formal dependability analysis using probability theory of HOL4. The proposed analysis provides the failure characteristics of the systems, i.e., reliability, availability, maintainability, etc. The authors formalized the Reliability Block Diagrams (RBD)~\cite{ahmad2014towards}, which are the graphical representations providing the functional behaviour of a system modules and their interconnections. The proposed formalization of RBD has been used for formally analyzing a simple oil and gas pipeline, a generic Virtual Data Center (VDC)~\cite{ahmed2016formalization}, Reliable Multi-Segment Transport (RMST) data transport, Event to Sink Reliable Transport (ESRT) protocols~\cite{ahmed2015formal} and Logistics Service Supply Chains (LSSCs)~\cite{ahmad2015towardslogistics}. Similarly, Ahmad et al.~\cite{ahmad2015towards} proposed a framework for the formal fault tree analysis using HOL4. The authors formalized the fault tree gates, i.e., AND, OR, NAND, NOR, XOR and NOT and formally verified their generic expressions for probabilities failures. Moreover, their proposed framework was used to perform the fault tree analysis of a solar array, which is used as a major source of power in the Dong Fang Hong-3 (DFH-3) satellite~\cite{ahmad2015towards} and a communication gateway software for the next generation Air Traffic Management System (ATMS)~\cite{ahmad2016formalization}.

Elderhalli et al.~\cite{elderhalli2018formal} developed a higher-order logic based framework for the formal dynamic dependability analysis using HOL4. The proposed analysis provides the dynamic failure characteristics of the systems, i.e., dynamic reliability and fault trees, etc. The authors formalized the Dynamic Fault Trees (DFTs)~\cite{elderhalli2019probabilistic} and Dynamic Reliability Block Diagrams (DRBD)~\cite{elderhalli2019formally} using HOL4. Moreover, they used their proposed formalization for formally analyzing the Drive-by-wire System (DBW), a Shuffle-exchange Network (SEN) and Cardiac Assist System (CAS)~\cite{elderhalli2019methodology}.

\section{Theorem Proving Support for CPS}

Table~\ref{TAB:libraries_for_formal_analysis} summarizes the formal libraries that are available in various theorem provers for performing the formal analysis of CPS. For example, the formal support for the dependability analysis of systems is only available in HOL4. Similarly, the libraries to formally reason about robotics and software components are available in most of the theorem provers. KeyMaera provides a support for formally analyzing the hybrid systems. Moreover, HOL4 and Isabelle theorem provers have a quite dense library for probabilistic and performance analyses of systems. Similarly, the transform methods are partially available in Isabelle, Coq and HOL4 theorem provers, i.e., only the Laplace transform is formalized in these theorem provers. However, HOL Light contains formal libraries for most of the transform methods, i.e., Laplace, Fourier and $z$-transforms. Also, the formal library for analyzing the optical systems is only available in HOL Light.

\begin{table}[]
\centering
\caption{Libraries for Formal Analysis in Major Theorem Provers}
\label{TAB:libraries_for_formal_analysis}
\begin{tabular}{|l|c|c|c|c|c|c|}
\hline
                     Analysis/Theorem Provers                                                    & HOL4 & HOL light & Isabelle/HOL & Coq & PVS & Keymaera \\ \hline
\begin{tabular}[c]{@{}l@{}}Transform Methods \end{tabular} &   \begin{normalsize} \cmark \end{normalsize}     &     \begin{normalsize} \cmark \end{normalsize}      &      \begin{normalsize} \cmark \end{normalsize}          &   \begin{normalsize} \cmark \end{normalsize}    &     &          \\ \hline
%%%%%%%%%%%%%%%%%%%%%%%%%%%%%%%%%%%%%%%%%%%%%%%%%%%%%%%%%%%%%%%%%%%%%%%%%%%%%%%%%%%%%%%%%%%%%%%%%%%%%%%%%%%%%%%%%%%%%%%%%%%%%%%%%%
\begin{tabular}[c]{@{}l@{}}Probabilistic Analysis\end{tabular}        &  \begin{normalsize} \cmark \end{normalsize}    &           &       \begin{normalsize} \cmark \end{normalsize}       &     &     &          \\ \hline
%%%%%%%%%%%%%%%%%%%%%%%%%%%%%%%%%%%%%%%%%%%%%%%%%%%%%%%%%%%%%%%%%%%%%%%%%%%%%%%%%%%%%%%%%%%%%%%%%%%%%%%%%%%%%%%%%%%%%%%%%%%%%%%%%%
\begin{tabular}[c]{@{}l@{}}Performance Analysis\end{tabular}         &   \begin{normalsize} \cmark \end{normalsize}   &           &         \begin{normalsize} \cmark \end{normalsize}     &     &     &          \\ \hline
%%%%%%%%%%%%%%%%%%%%%%%%%%%%%%%%%%%%%%%%%%%%%%%%%%%%%%%%%%%%%%%%%%%%%%%%%%%%%%%%%%%%%%%%%%%%%%%%%%%%%%%%%%%%%%%%%%%%%%%%%%%%%%%%%%
\begin{tabular}[c]{@{}l@{}}Dependability Analysis\end{tabular}         &  \begin{normalsize} \cmark \end{normalsize}    &           &              &     &     &          \\ \hline
%%%%%%%%%%%%%%%%%%%%%%%%%%%%%%%%%%%%%%%%%%%%%%%%%%%%%%%%%%%%%%%%%%%%%%%%%%%%%%%%%%%%%%%%%%%%%%%%%%%%%%%%%%%%%%%%%%%%%%%%%%%%%%%%%%
\begin{tabular}[c]{@{}l@{}}Hybrid Systems \end{tabular}         &      &           &              &     &     &     \begin{normalsize} \cmark \end{normalsize}     \\ \hline
%%%%%%%%%%%%%%%%%%%%%%%%%%%%%%%%%%%%%%%%%%%%%%%%%%%%%%%%%%%%%%%%%%%%%%%%%%%%%%%%%%%%%%%%%%%%%%%%%%%%%%%%%%%%%%%%%%%%%%%%%%%%%%%%%%
\begin{tabular}[c]{@{}l@{}}Optical Systems \end{tabular}         &      &     \begin{normalsize} \cmark \end{normalsize}      &              &     &     &          \\ \hline
%%%%%%%%%%%%%%%%%%%%%%%%%%%%%%%%%%%%%%%%%%%%%%%%%%%%%%%%%%%%%%%%%%%%%%%%%%%%%%%%%%%%%%%%%%%%%%%%%%%%%%%%%%%%%%%%%%%%%%%%%%%%%%%%%%
\begin{tabular}[c]{@{}l@{}}Quantum Systems \end{tabular}         &      &     \begin{normalsize} \cmark \end{normalsize}      &       \begin{normalsize} \cmark \end{normalsize}         &   \begin{normalsize} \cmark \end{normalsize}    &     &          \\ \hline
%%%%%%%%%%%%%%%%%%%%%%%%%%%%%%%%%%%%%%%%%%%%%%%%%%%%%%%%%%%%%%%%%%%%%%%%%%%%%%%%%%%%%%%%%%%%%%%%%%%%%%%%%%%%%%%%%%%%%%%%%%%%%%%%%%
\begin{tabular}[c]{@{}l@{}}Robotic Systems \end{tabular}               &   \begin{normalsize} \cmark \end{normalsize}   &   \begin{normalsize} \cmark \end{normalsize}          &       \begin{normalsize} \cmark \end{normalsize}       &     &   \begin{normalsize} \cmark \end{normalsize}  &    \begin{normalsize} \cmark \end{normalsize}      \\ \hline
%%%%%%%%%%%%%%%%%%%%%%%%%%%%%%%%%%%%%%%%%%%%%%%%%%%%%%%%%%%%%%%%%%%%%%%%%%%%%%%%%%%%%%%%%%%%%%%%%%%%%%%%%%%%%%%%%%%%%%%%%%%%%%%%%%
\begin{tabular}[c]{@{}l@{}}Software Components \end{tabular}         &   \begin{normalsize} \cmark \end{normalsize}     &     \begin{normalsize} \cmark \end{normalsize}        &    \begin{normalsize} \cmark \end{normalsize}          &     &   \begin{normalsize} \cmark \end{normalsize}    &    \begin{normalsize} \cmark \end{normalsize}        \\ \hline
%%%%%%%%%%%%%%%%%%%%%%%%%%%%%%%%%%%%%%%%%%%%%%%%%%%%%%%%%%%%%%%%%%%%%%%%%%%%%%%%%%%%%%%%%%%%%%%%%%%%%%%%%%%%%%%%%%%%%%%%%%%%%%%%%%
\end{tabular}
\end{table}

\section{Conclusion}

CPS are highly complex systems composed of actuators, sensors, and several electronic, communication and controller modules, and exhibit both the continuous and discrete dynamics. Due to the safety critical-nature of CPS, their accurate analysis is of utmost importance. This paper surveys some of the efforts that have been done regarding the formal verification of CPS using theorem proving by highlighting the aspects of CPS that have been verified using different theorem provers. In this regard, only one dedicated theorem prover, KeYmaera, has been developed for analyzing hybrid systems. However, we need to develop dedicated formal libraries in other theorem provers that can support the analysis of hybrid systems, i.e., incorporating the interlinked discrete and continuous-time features of a CPS simultaneously.

\bibliographystyle{splncs03}
\bibliography{bibliotex}

\end{document}